\documentclass[11pt]{article}
\usepackage{graphicx}
\usepackage{amsmath}
\usepackage{amssymb}
\usepackage{amsxtra}

\title{Charge asymmetry of new stable quarks in baryon asymmetrical Universe}
\author{Arnab Chaudhuri\\Department of Physics and Astronomy, Novosibirsk State University,\\ email: arnabchaudhuri.7@gmail.com\\Maxim Yu. Khlopov\\Institute of Physics, Southern Federal University,\\ Université de Paris, CNRS, Astroparticule et Cosmologie and\\National Research Nuclear University "MEPHI" \\ email: khlopov@apc.in2p3.f }

\begin{document}
\maketitle

\begin{abstract}
Effects of electroweak phase transition (EWPT) in balance
between baryon excess and the excess of stable quarks of new generation
is studied. With the conservation of SU(2) symmetry and other quantum
numbers, it makes possible sphaleron transitions between baryons,
leptons and new of leptons and quarks. A definite relationship between
the excess relative to baryon asymmetry is established. In passing by we
also show the small, yet negligible dilution in the pre-existing dark
matter density due the sphaleron transition.
\end{abstract}


\section{Introduction}\label{s:intro}
The matter-antimatter asymmetry, otherwise known as the baryon asymmetry of the universe (BAU) has been the focus of physicists for many a decade ~~\cite{RevModPhys.53.1,cohen,cohen2}. Various models have been developed to answer the question, ranging from the grand unified theory (GUT) to electroweak phase transition (EWPT). Irrespective of the mechanisms, the preexisting asymmetry is diluted by the baryon number violating mechanisms in the electroweak theory. This is due to the violation of the baryon and lepton number and the non-trivial topological structure
of the Yang–Mills theory.

The possibility to convert baryons into anti-leptons and the reverse exists in electroweak theory. The difference between the baryon and lepton numbers $(B-L)$ is conserved, even though individually, the quantum numbers are violated. Hence, it is important to know about the transition rates of such processes.

Sphalerons are generally associated with saddle points \cite{KM}, and is interpreted as the peak energy configuration, thus the transitions between vacua are associated with a violation of Baryon (and lepton) number. EWPT can be of first order, second order or a smooth crossover. Within the framework of the SM, it is a smooth crossover. However, BSM physics can lead to any of the three. The order of the phase transition can affect the outcome of the process. Entropy production and, in return, the dilution of preexisting frozen out species and baryon asymmetry can be some of them. Although baryon excess can be created at the time of electroweak symmetry breaking, it is preserved during the first order phase transition. In the second order, sphalerons can wipe out the total asymmetry created, but in first order, only the asymmetry created in the unbroken phase is wiped out. 

A recent overview of physics beyond the standard model and its cosmological signatures can be found in \cite{MK}, where it was shown that from the lack of supersymmetric particles at the LHC and from the positive results of the directly searched Weakly Interacting Massive Particles (WIMPs), the list of dark matter candidates can be strongly extended. The model of dark atoms of dark matter deserves special attention in this list, in light of its possibility to propose a nontrivial solution for the puzzles of direct dark matter searches, explaining the positive results of the DAMA/LIBRA experiment by the annual modulation of the low energy binding of dark atoms with sodium nuclei, which can be elusive in other experiments for direct WIMP searches. 

In this  approach, dark atoms represent a specific form of asymmetric, strongly interacting dark matter, being an atom-like state of stable $-2$ (or $-2n$) charged particles of a new origin bounded by a Coulomb interaction with (correspondingly, $n$) nuclei of primordial helium (see \cite{kuksa} for recent review and references). This explanation implies the development of a correct quantum mechanical description of the dark atom interaction with nuclei, which is now under way \cite{timur}.

Even though there are several models predicting $\pm 2$ ($\pm 2n$)-charged stable species, \cite{12,13,14,15,16,17,18,19,20,21,22,Norma,Norma2}), in this work we restrict our self to the 4th generation family as an extension to the standard model (SM) and proceed to study the electroweak phase transition (EWPT). The simplest charge-neutral model is considered here; also, we consider that EWPT is of the second order. In passing by, we show the dilution of pre-existing frozen out dark matter density in the presence of the 4th generation. 

The paper is organized as follows: In the next section, we talk about the 4th generation family, defining a definite relationship between the value and sign of the 4th generation family excess, relative to the baryon asymmetry, {which is due to the electroweak phase transition and possible sphaleron production being established}. 
 This is followed by the calculation of the dilution factor of pre-existing dark matter density, followed by a general conclusion.
 
\section{A brief review of $4^{th}$ generation}
The fourth generation is of theoretical interest in the context of sphaleron transition, electroweak symmetry breaking and large CP violating processes in the $4 \times 4$ CKM matrix, which may play a crucial role in understanding the baryon asymmetry in the universe. Thus, there are significant ongoing efforts to search for the fourth generation. In this work, we consider the stable 4th generation, which is basically constrained by contributions of virtual 4th generation particles in the Higgs boson decay rates, in the precision tests of standard model parameters, as well as by the LHC searches for R-hadrons, which mimic stable 4th generation stable hadrons. These constraints can still leave some room for the existence of such a family and an explanation of the puzzles of direct dark matter searches by dark atoms formed with primordial helium by $(\bar{U}\bar{U}\bar{U})$ antiquark clusters. 

Due to the excess of $\bar{{U}}$, only $-2$ charge or neutral 4th generation species are present in the universe. Indeed, stable antiquarks can form a $(\bar{U}\bar{U}\bar{U})$ cluster and a small fraction of neutral $\bar{U}u$  with ordinary $u$-quark. In principle, $(\bar{U}\bar{U}\bar{u})$ baryon should also be stable, but in a baryon asymmetrical universe, its interaction with ordinary baryons leads to its destruction in two $\bar{U}u$ mesons.  $^4He$, formed during the Big Bang nucleosynthesis, completely screens $Q^{--}$ charged hadrons in composite $[^4HeQ^{--}]$ ``atoms''. If this 4th family follows from string phenomenology, we have new charge ($F$) associated with 4th family fermions. Principally, $F$ should be the only conserved quantity but to keep matters simple, an analogy with WTC model is made and we assume two numbers: $FB$ (for 4th quark) and $L'$ (for 4th neutrino). Detailed calculations of WTC were made in
 \cite{12,18} and most of the terminology were kept the same as the above mentioned papers.
 
 As the universe expanded and the temperature decreased and the quantum number violating processes ceased to exist, the relation among the particles emerging from the process (SM + 4th generation) followed the following expression:
\begin{equation} \label{5}
   3(\mu_{u_L}+\mu_{d_L})+\mu+\mu_{U_L}+\mu_{D_L}+\mu_{L'}=0. 
\end{equation}
here, $\mu$ is the chemical potential of all the SM particles, $\mu_{L'}$ is the chemical potential of the new species leptons and $\mu_{U_L}$ and $\mu_{D_L}$ are that of the 4th generation quarks; see~\cite{18}. The number densities follow, respectively, for bosons and fermions:
\begin{equation} \label{6}
    n=g_*T^3\frac{\mu}{T}f(\frac{m}{T}),
\end{equation}
and
\begin{equation} \label{7}
    n=g_*T^3\frac{\mu}{T}g(\frac{m}{T}),
\end{equation}
where $f$ and $g$ are hyperbolic mathematical functions and $g_*$ is the effective degrees of freedom, which are given by the following:
\begin{equation} \label{7a}
    f(z)=\frac{1}{4\pi^2}\int_0^{\infty}x^2 cosh^{-2}\left(\frac{1}{2}\sqrt{z^2+x^2} \right)dx,
\end{equation}
and
\begin{equation} \label{7b}
    g(z)=\frac{1}{4\pi^2}\int_0^{\infty} x^2 sinh^{-2}\left(\frac{1}{2}\sqrt{z^2+x^2} \right)dx.
\end{equation}

The number density of baryons follows the following expression:
\begin{equation} \label{8}
    B=\frac{n_B-n_{\bar{B}}}{gT^2/6}.
\end{equation}

As the main point of interest is the ratio of baryon excess to the excess of the stable 4th generation, the normalization cancels out, without loss of generality. 

Let us define a parameter $\sigma$, which, respectively, for fermions and bosons are given by the following:
\begin{equation} \label{9}
    \sigma=6f\frac{m}{T_c},
\end{equation}
and
\begin{equation} \label{10}
    \sigma=6g\frac{m}{T_c}.
\end{equation}

$T_c$ is the transition temperature and is given by the following:
\begin{equation} \label{10a}
    T_c=\frac{2M_W(T_c)}{\alpha_W~ln(M_{pl}/T_c)}B(\frac{\lambda}{\alpha_W}).
\end{equation}

In the above equation, $M_W$ is the mass of W-boson, $M_{pl}$ is the Planck mass and $\lambda$ is the self-coupling. The function $B$ is derived from experiment and takes the value from 1.5 to 2.7.

The new generation charge is calculated to be the following:
\begin{equation} \label{11}
    FB=\frac{2}{3}(\sigma_{U_L}\mu_{U_L}+\sigma_{U_L}\mu_{D_L}+\sigma_{D_L}\mu_{D_L}),
\end{equation}
where $FB$ corresponds to the anti-U ($\bar{U}$) excess. For detailed calculations, please see \cite{18}. 

The SM baryonic and leptonic quantum numbers are expressed as  the following:
\begin{equation} \label{10a1}
    B=\left[(2+\sigma_t)(\mu_{uL}+\mu_{uR})+3(\mu_{dL}+\mu_{dR}) \right]
\end{equation}
and
\begin{equation} \label{10b}
    L=4\mu+6\mu_W
\end{equation}
where in Equation (\ref{10a1}), the factor 3 of down-type
quarks is the number of families. For the 4th generation lepton family, the quantum number is given by the following:
\begin{equation} \label{10c}
    L'=2(\sigma_{\nu'}+\sigma_{U_L})\mu_{\nu'L}+2\sigma_{U_L}\mu_W+(\sigma_{\nu'}-\sigma_{U_L})\mu_0
\end{equation}
where $\nu'$ is the new family of neutrinos originating from the extension of SM, and $\mu_0$ is the chemical potential from the Higgs sector in SM.

Due to the presence of a single Higgs particle, the phase transition is of the second order. The ratio of the number densities of the 4th generation to the baryons is determined by the following:
\begin{equation} \label{12}
    \frac{\Omega_{FB}}{\Omega_B}=\frac{3}{2}\frac{FB}{B}\frac{m_{FB}}{m_p}.
\end{equation}

The electrical neutrality and negligibly small chemical potential of the Higgs sector is the result of the second order phase transition. The ratio of the number density of the 4th generation to the baryon number density can be expressed as a function of the ratio of the original and new quantum numbers. In the limiting case of the second order EWPT, we obtain the following:
\begin{equation} \label{13}
    -\frac{FB}{B}=\frac{\sigma_{U_L}}{3(18+\sigma_{\nu'})}\left[(17+\sigma_{\nu'})+\frac{(21+\sigma_{\nu'})}{3}\frac{L}{B}+\frac{2}{3}\frac{9+5\sigma_{\nu'}}{\sigma_{\nu'}}\frac{L'}{B}\right].
\end{equation}

In the following Figure \ref{f-dm}, the predicted relationship between the frozen out excess of $\bar{U}$-antiquarks and baryon asymmetry is shown as a function of U-quark mass $m$. The minimal mass $m$  can be determined from the R-hadrons search at the LHC as $1$ TeV. The predicted contribution of dark atoms, in which ($\bar{U}\bar{U}\bar{U}$) are bound with primordial helium nuclei, can explain the observed dark matter density at $m \sim 3.5$ TeV, which is compatible with the above mentioned experimental lower limit.

\begin{figure}[h!]
\includegraphics[]{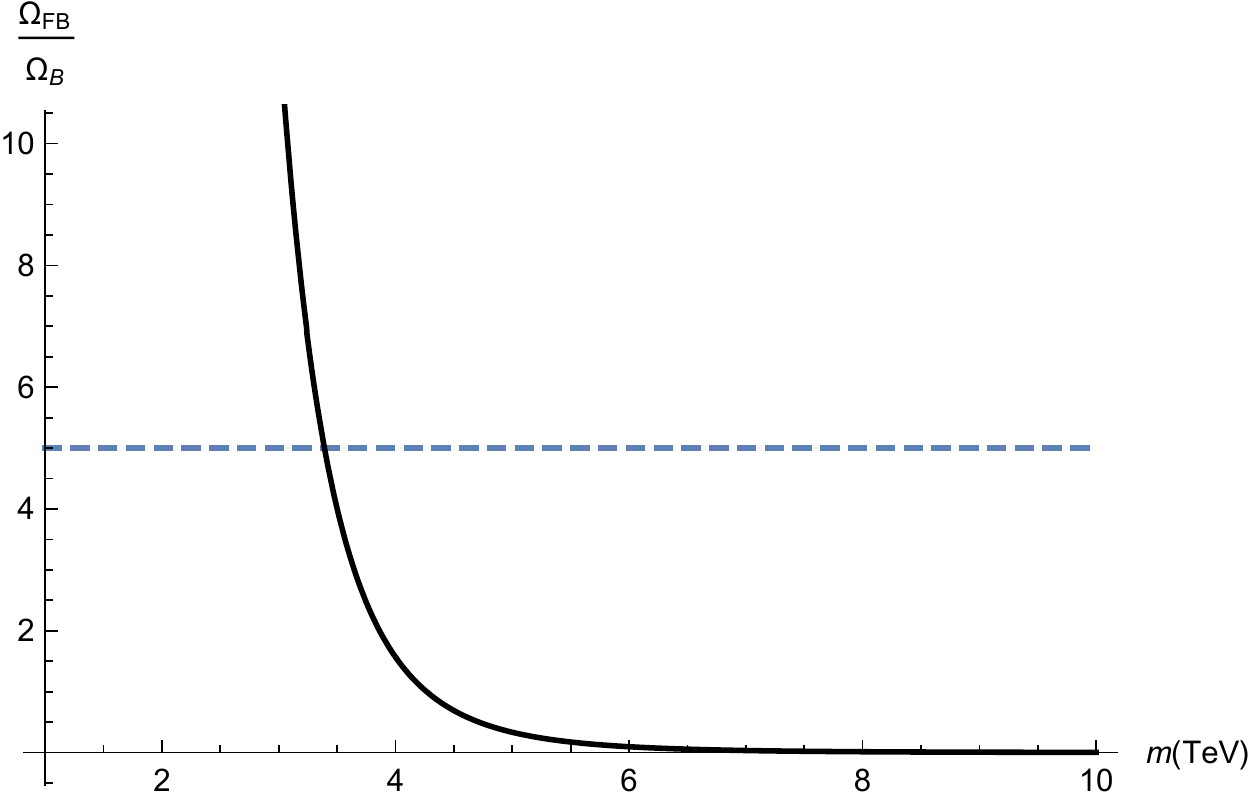}
\caption{The ratio of dark matter and baryon densities as a function of the U-quark mass ($m$). This ratio is frozen out at the critical temperature of the assumed second order EWPT $T=T_c=179$ GeV.  At the U-quark mass $m \approx 3.5$ TeV, the predicted density of dark atoms, formed by ($\bar{U}\bar{U}\bar{U}$) bound with primordial helium nuclei, can explain the observed dark matter density.}
\label{f-dm}
\end{figure} 

Hence, we establish a relationship between the baryon excess and the excess of $\bar{U}$ for the second order EWPT.

\section{Dilution of Pre-Existing Dark Matter Density}
The thermodynamic quantity, entropy density, is a conserved quantity in the initial stage of universe expansion, especially when the primeval plasma is in thermal equilibrium with a negligible chemical potential. As soon as the universe enters into the state of thermal non-equilibrium, i.e., when $\Gamma>H$, where $\Gamma$ is the reaction rate and $H$ is the Hubble parameter, the conservation law breaks down, and entropy starts pouring into the plasma; this can dilute the pre-existing baryon asymmetry and dark matter density. 

There are many instances of entropy production, such as primordial black hole evaporation~\cite{Chaudhuri:2020wjo}, electroweak phase transition within the standard model and the two Higgs doublet model~\cite{Chaudhuri:2017icn,Chaudhuri:2021agl,AMS}. Apart from these, the freeze out of dark matter density might lead to entropy production, which in turn, can dilute the pre-existing dark matter density. 

The Lagrangian theory consists of the Langrangian of the standard model (SM) and the interaction terms of the 4th generation fermionic family. It is given by the following:
\begin{equation} \label{lag}
    \mathcal{L}=\mathcal{L}_{SM}+\mathcal{L}_{4^{th}}, 
\end{equation}
where $\mathcal{L}_{SM}$ is given by the following:
\begin{equation}
    \mathcal{L}_{SM}=\frac{1}{2}g^{\mu \nu}\partial_{\mu}\phi \partial_{\nu}\phi -U_{\phi}(\phi)+\sum_{j} i\left[g^{\mu \nu} \partial_{\mu} \chi_j^{\dagger} \partial_{\nu} \chi_j - U_j(\chi_j) \right].
\end{equation}

The CP violating potential of the theory is as follows:
\begin{equation}
    U_{\phi}(\phi)=\frac{\lambda}{4}\left(\phi^2-\eta^2 \right)^2+\frac{T^2 \phi^2}{2}\sum_j h_j \left(\frac{m_j(T)}{T} \right).
\end{equation}
here, $\lambda=0.13$ is the quartic coupling constant and $\eta$ is the vacuum expectation values, which is $\sim$246~GeV in the SM. $T$ is the plasma temperature and $m_j(T)$ is the mass of the $\chi_j$-particle at temperature \emph{T}; see~\cite{imelo}.

To calculate the dilution factor, it is necessary to compute the energy and the pressure density of the plasma, using the energy--momentum tensor. Following the detailed calculation in~\cite{Chaudhuri:2021agl} and assuming that the universe was flat in the early epoch with the metric $g_{\mu \nu}=(+,-,-,-)$, we have the following:
\begin{equation} \label{p+r}
    \rho+\mathcal{P}=\dot{\phi}^2+\frac{4}{3}\frac{\pi^2 g_*}{30}T^4.
\end{equation}

In order to proceed with the calculation of the dilution factor, the EWPT transition temperature needs to be calculated first. The transition temperature is derived using the following expression:
\begin{equation} \label{Tc}
    V(\phi=0,T=T_c)=V(\phi=\eta, T=T_c).
\end{equation}
here, $\eta$ is the vacuum expectation value and $T_c$ is the transition temperature. In Equation~(\ref{Tc}), substituting the values of the standard model particles and the minimal allowed masses of the 4th generation particle, which can be estimated from the R-hadrons search at the LHC as 1 TeV, $T_c$ is found to be $\sim$179~GeV. With a range of allowed values, one can obtain a range of $T_c$s and study the nature of the EWPT and other related properties, but that is beyond the scope of the current paper. With proper data and tools, this analysis will certainly be made in the near future.

The last term in Equation (\ref{p+r}) arises from the Yukawa interaction between fermions. The Higgs field starts to oscillate around the minimum, which appears during the phase transition. Particle production from this oscillating field causes damping. The characteristic time of decay is equal to the decay width of the Higgs bosons. If it is large in comparison to the expansion, and thus the universe cooling rate, then we may assume that the Higgs bosons essentially live in the minimum of the potential. This was clearly discussed in~\cite{Chaudhuri:2017icn}.

To calculate the entropy production, it is necessary to solve the evolution equation for energy density conservation as follows: 
\begin{equation} \label{fried}
\Dot{\rho}=-3H(\rho+P).
\end{equation}

\begin{figure}[h!]
\includegraphics[]{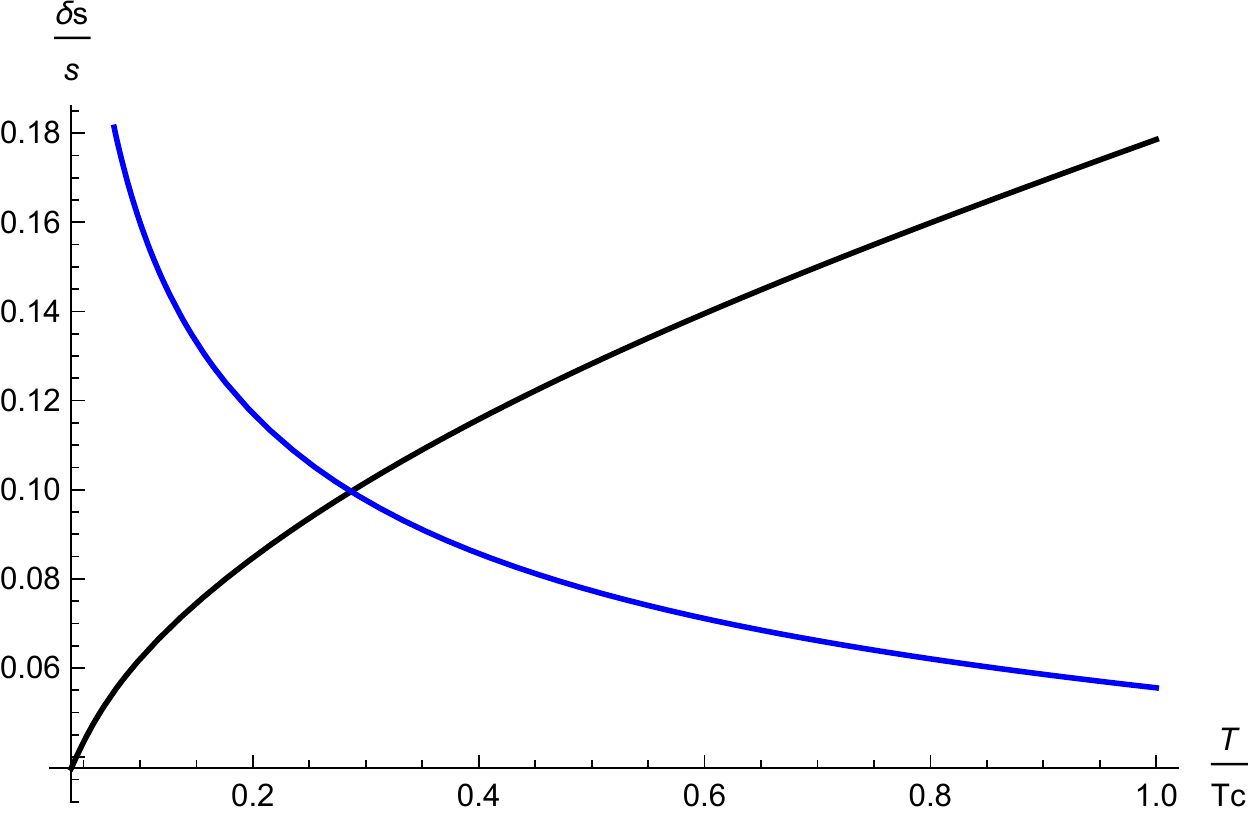}
\caption{Entropy 
production (black line) and the dilution of pre-existing dark matter (blue line) in the presence of 4th generation fermions are presented.  }
\label{f-entropy-1}
\end{figure}

In Figure \ref{f-entropy-1}, both the dilution of the pre-existing dark matter (blue line) and the entropy production in the presence of 4th generation lepton family (black line) are shown. It is clear that since the sphaleron transition is of the second order, the net dilution and entropy production $(\sim$$18\%)$ are somewhat low compared to the scenarios of the first order. Again, the presence of a single Higgs field makes the phase transition of the second order.

\section{Conclusions}

In the present paper, we have deduced a definite relationship between the value and sign of the 4th generation family excess and baryon asymmetry due to the sphaleron effects frozen out at the electroweak phase transition, as is clear from Equation (\ref{13}) and Figure \ref{f-dm}.

At the transition temperature, $T_c=179$~GeV and the mass of the stable $U$-quark of the new family $m \approx 3.5$~TeV, the predicted density of the dark atoms can explain the observed cosmological dark matter density.
This value and experimental constraints on the contribution of new electroweakly interacting fermions appeal to the involvement of additional Higgs bosons, whose existence can influence the value of $T_c$ and, correspondingly, the determination of the mass of the $U$-quark, for which dark atoms explain the dark matter density. Being beyond the scope of the present work, such a self-consistent analysis of the models with new stable quarks, accompanied by an extended Higgs sector, can open up a new specific direction of studies of BSM physics.

The search for new physics and dark matter has been an ongoing area of research for decades. Even though there are many multi-Higgs models, there are few multi-charged models present. The theory of the 4th generation can serve to leap toward new physics in the framework of heterotic string phenomenology.  As seen from the work, just like the standard model, we can link the stable quarks of the new generation particles with the baryon asymmetry theoretically. The existence of new stable quarks with the SM electroweak charges can follow from other unifying schemes (in the approach \cite{Norma,Norma2} in particular); the important conclusion of our work is that balancing baryon asymmetry with sphaleron transitions can provide an excess of $\bar{U}$ antiquarks, forming a $(\bar{U}\bar{U}\bar{U})$ `core' of dark atoms in which it is bound by a Coulomb force with primordial helium. The possibility of dark atoms extends the list of possible dark matter candidates, predicted in such models. To make new quarks with electroweak charges compatible with the data on the Higgs boson decay rates, their coupling to the SM Higgs boson should be suppressed, and they should acquire their mass from coupling to other Higgs bosons \cite{shib}. It would imply the accomplishment of models with new stable generations with SM electroweak charges by multi-Higgs models, opening up a probe of studying the Higgs and electroweak symmetry breaking sectors in a rigorous manner.

Dark matter candidates in the form of bounded dark atom can emerge from this model, due to the excess of $\bar{U}$ within the primordial He nuclei. We have considered only the lightest and most stable particles and also took into account only the second order phase transition. The dilution of pre-existing dark matter density was calculated; in the present scenarios, the dark matter density was reduced by $\sim$$18\%$.

\section*{Acknowledgements}
The work of A.C. is funded by RSF Grant 19-42-02004. The research by M.K. was supported by the Ministry of Science and Higher Education of the Russian Federation under Project "Fundamental problems of cosmic rays and dark matter", No. 0723-2020-0040.



\begin{thebibliography}{99}
\bibitem{RevModPhys.53.1}
Dolgov, A.D.; Zeldovich, Y.B. Cosmology and elementary particles. {\em Rev. Mod. Phys.} {\bf 1981}, {\em 53}, 1, doi:10.1103/RevModPhys.53.1.
\bibitem{cohen}
Cohen, A.G.; Kaplan, D.B.; Nelson, A.E. Progress in Electroweak Baryogenesis. {\em Ann. Rev. Nucl. Part. Sci.} {\bf 1993}, {\em43}, 27--70, doi:10.1146/annurev.ns.43.120193.000331.

\bibitem{cohen2}
Cohen,A.G.; Kaplan, D.B.; Nelson, A.E. Diffusion enhances spontaneous electroweak baryogenesis. {\em Phys. Lett. B} {\bf 1994}, \emph{336}, 41--47, doi:10.1016/0370-2693(94)00935-X.

\bibitem{KM}
Klinkhamer, F.R.; Manton, N.S. A saddle-point
solution in the Weinberg-Salam theory. {\em Phys. Rev. D} {\bf 1984}, {\em 30}, 2212–2220.

\bibitem{MK}
Khlopov. M.Y. What comes after the Standard Model? {\em Prog. Part. Nucl. Phys.} {\bf 2021}, {\em 116},  103824, doi:10.1016/j.ppnp.2020.103824

\bibitem{kuksa} Beylin, V.; Khlopov, M.; Kuksa, V.;  Volchaanskiy, N.  New physics of strong interaction and Dark Universe. {\em Universe} {\bf 2020}, {\em 6}, 196.

\bibitem{timur}  Bikbaev, T.E.; Khlopov, M.Y.; Mayorov, A.G.  Numerical simulation of dark atom interaction with nuclei. {\em Bled Work. Phys.} \mbox{{\bf 2020}, {\em 21}, 105--117}.

\bibitem{12}
Khlopov, M.Y.; Kouvaris, C. Strong interactive massive particles from a strong coupled theory. {\em Phys. Rev. D} {\bf 2008}, {\em77}, 065002, doi:10.1103/PhysRevD.77.065002.

\bibitem{13}
Sannino, F.; Tuominen, K.
Orientifold theory dynamics and symmetry breaking. {\em Phys. Rev. D} {\bf 2005}, {\em 71}, 051901, doi:10.1103/PhysRevD.71.051901.

\bibitem{14}
Hong, D.K.;  Hsu, S.D.H.; Sannino, F. Composite Higgs from higher representations. {\em Phys. Lett. B} {\bf 2004},  {\em 597}, 89, doi:10.1016/j.physletb.2004.07.007.

\bibitem{15}
Dietrich, D.D.; Sannino, F.; Tuominen, K. Light composite Higgs boson from higher representations versus electroweak precision measurements: predictions for CERN LHC. {\em Phys. Rev. D} {\bf 2005}, {\em 72}, 055001, doi:10.1103/PhysRevD.72.055001.

\bibitem{16}
Dietrich, D.D.; Sannino, F.;  Tuominen, K. Light composite Higgs and precision electroweak measurements on the 
Z resonance: An update. {\em Phys. Rev. D} {\bf 2006}, \emph{73},  037701, doi:10.1103/PhysRevD.73.037701.


\bibitem{17}
Gudnason, S.B.; Kouvaris, C.; Sannino F. Towards working technicolor: Effective theories and dark matter. {\em Phys. Rev. D} \mbox{{\bf 2006}, {\em 73}, 115003}, doi:10.1103/PhysRevD.73.115003.

\bibitem{18}
{Gudnason, S.B.; Kouvaris, C.; Sannino, F. Dark matter from new technicolor theories.} 
 {\em Phys. Rev. D} {\bf 2006}, {\em 74}, 095008, doi:10.1103/PhysRevD.74.095008.

\bibitem{19}
Belotsky, K.M.; Khlopov, M.Y.; Shibaev, K.I. Stable quarks of the 4th family? In \emph{Physics of Quarks: New Research;  Horizons in World Physics}

\bibitem{20}
Khlopov, M.Y. Composite dark matter from the fourth generation. {\em JETP Lett.}  {\bf 2006}, {\em 83}, 1, doi:10.1134/S0021364006010012 .

\bibitem{21}
Belotsky, K.M.; Khlopov, M.Y.; Shibaev, K.I. Stable matter of 4th generation: Hidden in the Universe and close to detection? \emph{arXiv} \textbf{2006},  arXiv:astro-ph/0602261.
 
 \bibitem{22}
Belotsky, K.M.; Khlopov, M.Y.; Shibaev, K.I. Composite Dark Matter and its Charged Constituents.  {\em Gravit. Cosmol.} {\bf 2006}, {\em 12}, 1--7, doi:10.1134/S0202289312020028.

\bibitem{Norma}
Mankoc-Borstnik, N. Unification of spins and charges in Grassmann space? {\em Mod. Phys. Lett.} {\bf 1995}, {\em 10}, 587--596.

\bibitem{Norma2} Mankoc Borstnik, N.S. The spin-charge-family theory is explaining the origin of families, of
the Higgs and the Yukawa couplings. \emph{J.  Mod. Phys. } \textbf{2013}, \emph{4},  823. 

\bibitem{Chaudhuri:2020wjo}
Chaudhuri, A.; Dolgov, A. PBH evaporation, baryon asymmetry and dark matter.
{\em arXiv} {\bf 2020}, arXiv:2001.11219.

\bibitem{Chaudhuri:2017icn}
Chaudhuri, A.; Dolgov, A.
Electroweak phase transition and entropy release in the early universe.
\emph{JCAP} {\bf 2018}, \emph{2018},  032, doi:10.1088/1475-7516/2018/01/032. 

\bibitem{Chaudhuri:2021agl}
Chaudhuri, A.; Khlopov, M.Y.
Entropy production due to electroweak phase transition in the framework of two Higgs doublet model. 
\emph{Physics} {\bf 2021},  {\em 3}, 275--289, doi:10.3390/physics3020020.

\bibitem{AMS}
Chaudhuri, A.; Khlopov, M.Y.; Porey, S. 
Effects of 2HDM in electroweak phase transition. \emph{Galaxies} {\bf 2021}, {\em 9}, 45, 
\linebreak doi:10.3390/galaxies9020045.

\bibitem{imelo}
 Melo, I. Higgs potential and fundamental physics. {\em Eur. J. Phys.} {\bf2017}, {\em 38}, 065404, doi:10.1088/1361-6404/aa8c3d.

\bibitem{shib}
Khlopov, M.Y.; Shibaev, R.M. Probes for 4th generation constituents of dark atoms in Higgs boson studies at the LHC.  \emph{Adv. High Energy Phys.} {\bf 2014}, \emph{2014},  406458. 

\end{thebibliography}
\end{document}